\documentclass[aps,prl,twocolumn]{revtex4-1}
\usepackage{hyperref}
\hypersetup{colorlinks=true, citecolor=blue, urlcolor=blue, linkcolor=blue}

\pdfoutput=1
\usepackage[utf8]{inputenc}
\usepackage[T1]{fontenc}
\usepackage{amssymb}
\usepackage{amsmath}
\usepackage{graphicx}
\usepackage{subcaption}
\usepackage{color}
\usepackage{calc}
\usepackage{mathtools}
\usepackage{tabularx}

\renewcommand{\vec}[1]{\mathbf{#1}}

\usepackage[normalem]{ulem}

\newcommand{\Eq}[1]{Eq. ~\eqref{#1}}
\newcommand{\Fig}[1]{Fig. ~\ref{#1}}

\begin{document}

\title{
Averting the Infrared Catastrophe in the Gold Standard of Quantum Chemistry
}
\author{Nikolaos Masios}
\author{Andreas Irmler}
\author{Tobias Sch\"afer}
\author{Andreas Gr\"uneis}
\email{andreas.grueneis@tuwien.ac.at}
\affiliation{
  Institute for Theoretical Physics, TU Wien,\\
  Wiedner Hauptstraße 8-10/136, 1040 Vienna, Austria
}
\date{\today, PREPRINT}


\begin{abstract}
Coupled-cluster theories can be used to compute \emph{ab initio} electronic correlation energies
of real materials with systematically improvable accuracy.
However, the widely used coupled cluster singles and doubles plus perturbative triples [CCSD(T)] method is only applicable to insulating materials.
For zero-gap materials the truncation of the underlying many-body perturbation expansion
leads to an infrared catastrophe.
Here, we present a novel
perturbative triples formalism that yields convergent correlation energies
in metallic systems. Furthermore, the computed correlation energies for the
three-dimensional uniform electron gas at metallic densities are in good agreement with
quantum Monte Carlo results. At the same time the newly proposed method
retains all desirable properties of CCSD(T) such as its accuracy for insulating systems
as well as its low computational cost compared to a full inclusion of the triples.
This paves the way for \emph{ab initio} calculations of real metals
with chemical accuracy.
\end{abstract}

\maketitle


\emph{Introduction.} ---- \emph{Ab initio} methods that achieve
systematically improvable accuracy
for metallic systems are urgently needed to understand chemical reactions on metal surfaces
or to compute the thermodynamic stability of materials.
Currently available exchange and correlation energy density functionals
often fail to achieve the desired
level of accuracy compared to experiment. A prominent failure includes the
incorrect prediction of molecular adsorption sites
on metal surfaces~\cite{Feibelman2001}.  
As an alternative, more accurate quantum Monte Carlo (QMC) calculations can be applied to
metals~\cite{Pozzo2008,Doblhoff2017}. 
However, even diffusion QMC calculations exhibit a strong dependence on the
fixed node approximation~\cite{Nemec2010}.
Compared to QMC and density functional theory, many-electron perturbation theories offer a conceptually different approach to
solve the many-electron problem with high accuracy.  In
particular, coupled-cluster (CC) theory offers a systematically improvable ansatz
for the many-electron wave function employing a series of
higher order particle-hole excitation operators.
While systems exhibiting strong correlation, e.g. stretched covalent bonds,
require high-order excitation operators or multireference approaches,
single reference systems are already described accurately using low orders~\cite{bartlett2007}.
In particular, at the truncation level of  single, double and
perturbative triple particle–hole excitation operators, CCSD(T) theory predicts
atomization and reaction energies for a wide class of single reference molecules with an
accuracy of approximately 1~kcal/mol~\cite{bartlett2007}.
As such, CCSD(T) is often referred to as the ``gold standard'' of molecular quantum chemistry.
This also motivated recent applications of coupled-cluster theory to study
solids~\cite{Booth2013,McClain2017,Yang640,grueneis2015b,Gruber2018,Weiler2022}, where
highly accurate predictions of, for example, pressure-temperature phase diagrams~\cite{gruber18b}
could be achieved.
However, it should be noted that such calculations require a careful convergence
with respect to the employed basis sets and system size, which 
is much more complicated than for lower levels of theory~\cite{McClain2017,Gruber2018,Irmler2021}.

Moreover, metallic systems still constitute a major challenge for currently available CC theories.
Although CCSD can be applied to metals, recently obtained results for a number
of metallic systems indicate that CCSD falls short of achieving chemical accuracy in metals,
which is expected and agrees with findings for molecules and insulating solids~\cite{Mihm2021,Neufeld2022}.
The inclusion of the perturbative triples correction (T) though,
is not possible due to the infrared catastrophe
caused by the truncation of the many-electron perturbation expansion~\cite{Shepherd2013}.
The infrared catastrophe leads to a divergence of the CCSD(T) correlation energy per electron for metals
for increasing simulation cell sizes also referred to as the thermodynamic limit (TDL).
A full nonperturbative inclusion of the triple excitation operator in the CC method
is convergent but computationally too expensive and can only be applied to few relatively small systems~\cite{Neufeld2017}.
Here, we present a modification to
the perturbative triples theory that is applicable to metals and
retains all desirable properties including its accuracy for insulating systems
and low computational cost compared to a full inclusion of the triples.

\emph{Theory.} ----
To better understand the infrared catastrophe of perturbation theories denoted as $(X)$,
we  make use of the following three quantities:
the correlation energy $E_\text{c}^{(X)}$, the electronic transition structure factor $S^{(X)}(\mathbf{q})$
and the quantity $T^{(X)}_i(\mathbf{q})$.
The correlation energy is defined as
\begin{equation}
\label{eq:ccdCorr}
        E_\text{c}^{(X)} = \sum_{\mathbf{q}}\upsilon(\mathbf{q})\underbrace{ \Biggl[\, 
               \frac{\delta\upsilon_{ij}^{ab}}{\delta\upsilon( \mathbf{q} )}
               \left(2 t_{ij}^{ab} - t_{ji}^{ab} \right)_{(X)} \Biggr] }_{\coloneqq S^{(X)}(\mathbf{q})}.
\end{equation}
%
%
%
The indices $i$, $j$ and $a$, $b$ denote occupied and virtual spatial
orbitals, respectively. Einstein summation convention applies to repeated indices throughout this work.
We will first focus on second-order perturbation theory,
direct ring CC doubles (rCCD) theory, which is closely related to the random-phase approximation (RPA), 
and CC doubles theory (CCD).
We note that, due to the symmetry of the  uniform electron gas (UEG) Hamiltonian, single excitations
are absent.
Furthermore, in the UEG, the one-electron orbitals are plane waves with wave vectors $\vec
k_i$, $\vec k_j$ and $\vec k_a$, $\vec k_b$.  This allows us to write the
two-electron repulsion integral as
\begin{math}\upsilon_{ij}^{ab} = \upsilon(\mathbf{q})
\delta_{\mathbf{k}_i-\mathbf{k}_a, \mathbf{k}_b - \mathbf{k}_j}\end{math},
with the momentum transfer vector ${\mathbf{q} = \mathbf{k}_i - \mathbf{k}_a}$,
${\upsilon(\mathbf{q}) = \frac{4 \pi}{\Omega \left| \mathbf{q} \right|^2}}$
and $\Omega$ being the volume of the simulation cell. 
The functional derivative
$  \frac{\delta \upsilon_{ij}^{ab}}{\delta \upsilon(\mathbf{q})} =\delta_{\mathbf{q}, \mathbf{k}_b - \mathbf{k}_j} \delta_{\mathbf{q}, \mathbf{k}_i - \mathbf{k}_a} $
enables a concise notation.
The amplitudes $t_{ij}^{ab}$ are obtained by solving the amplitude equations of the employed many-electron
perturbation theory.

\Eq{eq:ccdCorr} introduces the electronic transition structure factor $S(\mathbf{q})$,
which gives access to the dependence of the
correlation energy on the interelectronic interaction distance.
An additional quantity of significance for the present work is given by
\begin{equation}
T^{(X)}_i(\mathbf{q})=
\Biggl[\, 
               \frac{\delta\upsilon_{ij}^{ab}}{\delta\upsilon( \mathbf{q} )}
               {t_{ij}^{ab}}_{(X)} \Biggr].
\end{equation}
Similar to the structure factor, $T^{(X)}_i(\mathbf{q})$ depends on the momentum transfer vector $\mathbf{q}$.

\emph{The infrared catastrophe.} ----
Having introduced the most important quantities needed for our analysis,
we turn to the case of second-order perturbation theory, which
is a textbook example for the infrared catastrophe. In particular,
we focus on second-order M\o ller-Plesset perturbation theory (MP2), which employs the Hartree-Fock (HF) Hamiltonian
as the unperturbed reference system~\cite{moller_note_1934}.

\begin{figure*}
  \centering
 \includegraphics[width=0.23\linewidth]{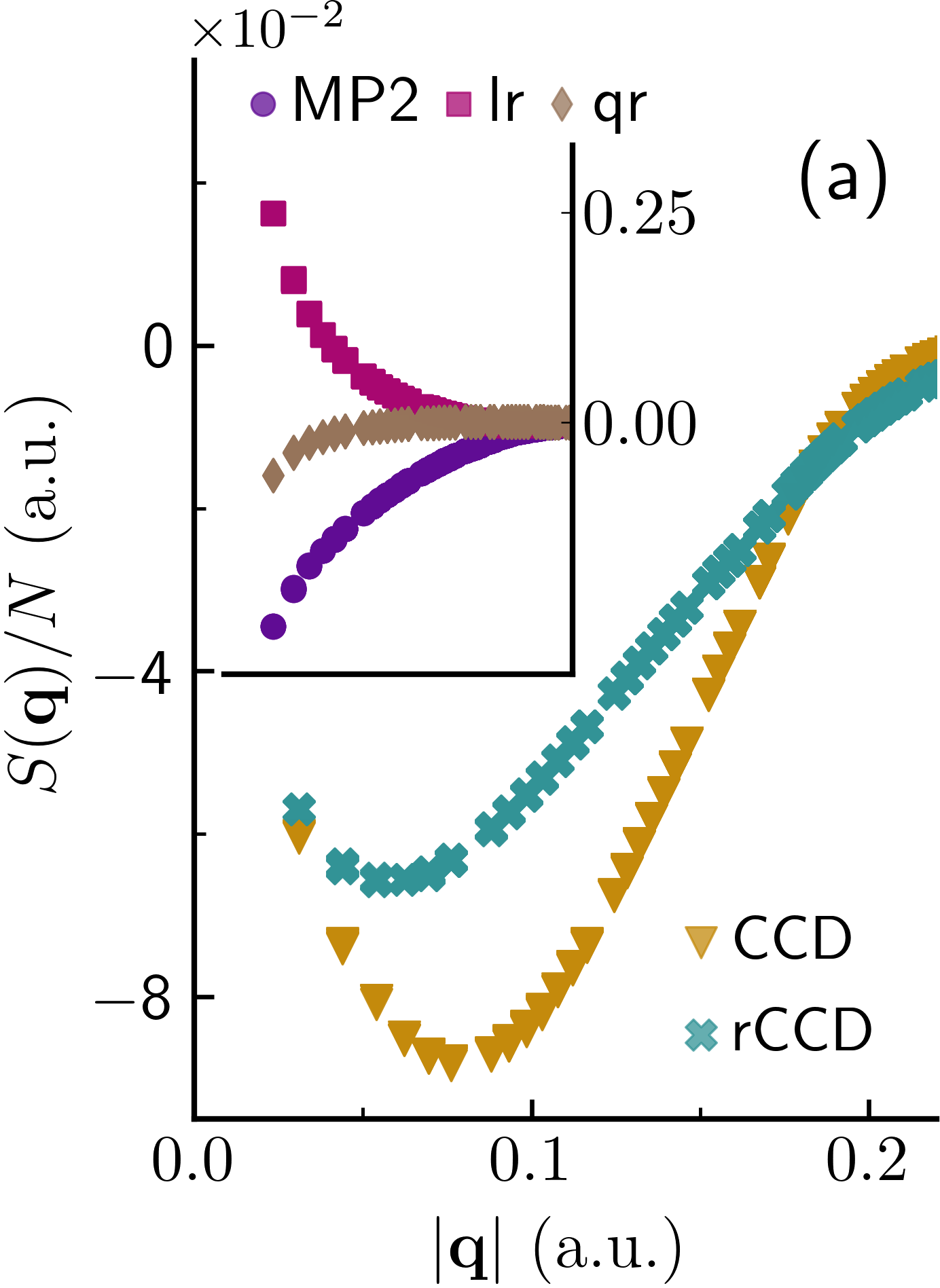}
 \includegraphics[width=0.21\linewidth]{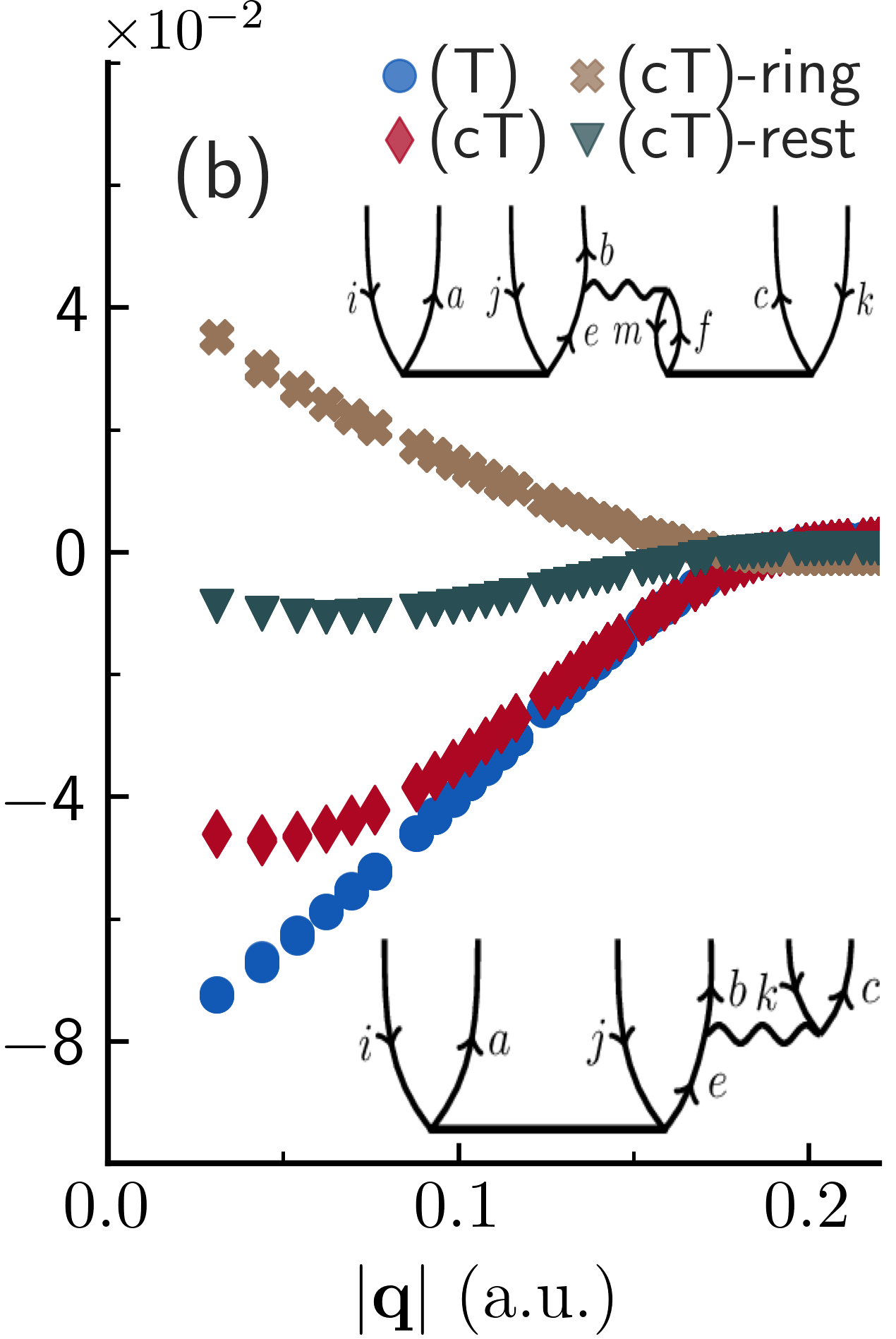}
 \includegraphics[width=0.24\linewidth]{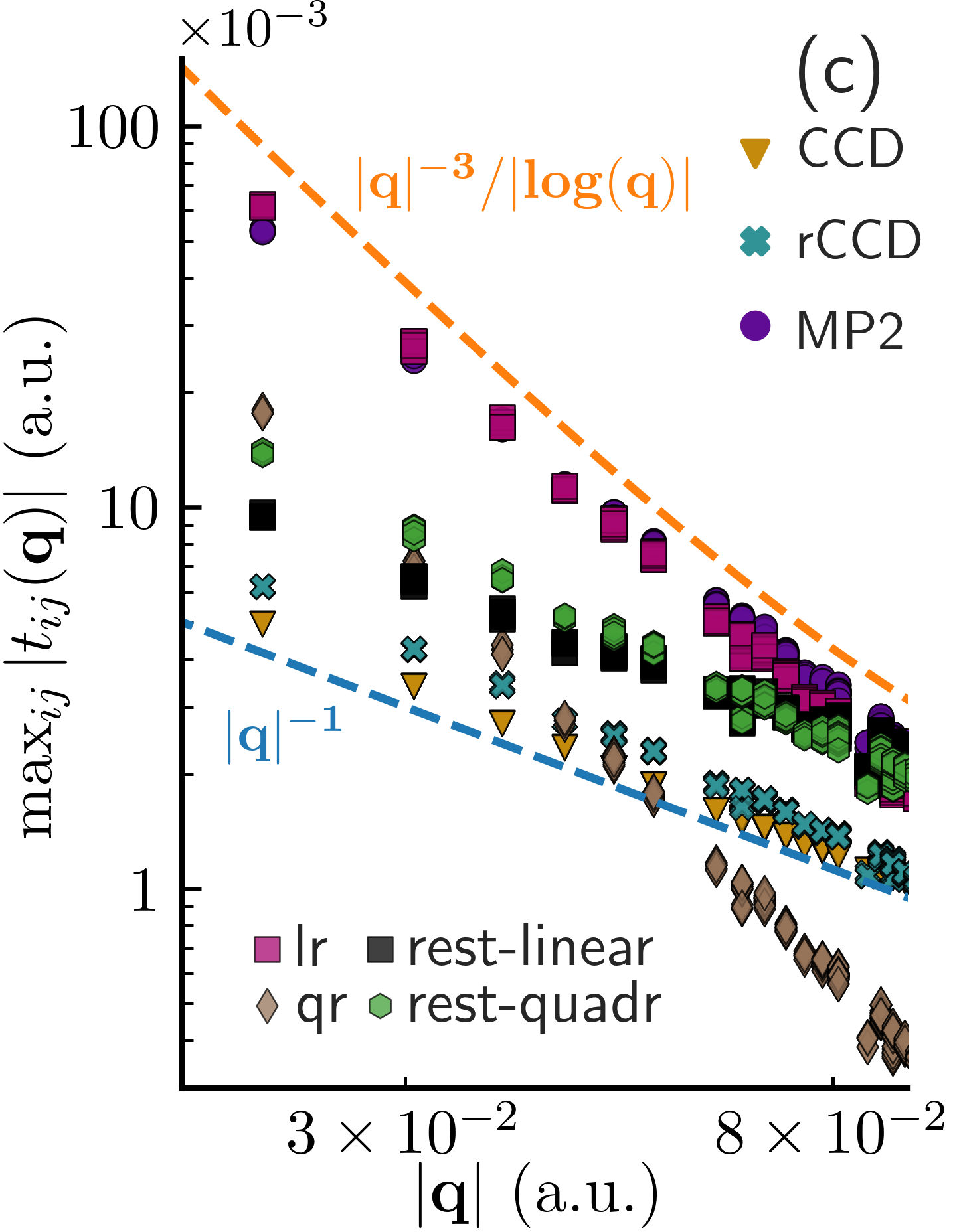}
 \includegraphics[width=0.24\linewidth]{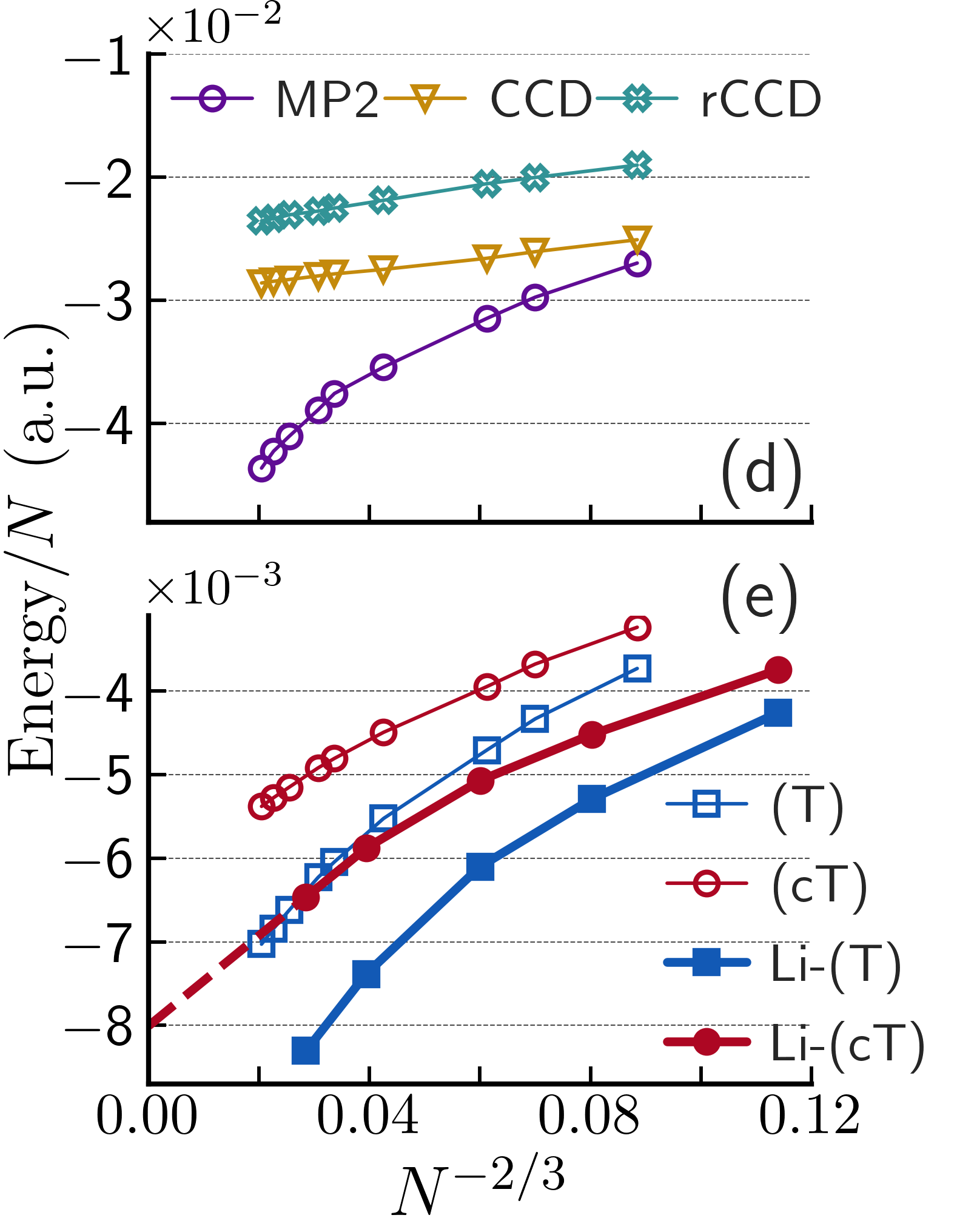}
  \caption {
           (a) and (b) UEG transition structure factor contributions at different levels of theory
           for 246 electrons with 2178 spatial orbitals, 266 twists for rCCD, 166 twists for CCD, (T) and (cT) and $r_{s}=20\,$a.u.
           (c) Contributions to $t_{ij}(\mathbf{q})$ at different levels of theory for 730 electrons with
           6254 spatial orbitals, 40 twists and $r_{s}=20$~a.u.
           (d) and (e) CBS correlation energies per electron with 40 twists at $r_s=3\,$a.u. for the UEG retrieved as $N^{-2/3}$, where $N$ is the electron number.
           (d) MP2, rCCD, and CCD correlation energies. (e) (T) and (cT) energies. 
           (e) (T) and (cT) correlation energy per electron with 20 twists for Li denoted as Li-(T) and Li-(cT). Details in the Supplemental Material.
           }
            \label{fig:sofg}
\end{figure*}

The MP2 correlation energy is given by Eq.~(\ref{eq:ccdCorr}) using $t_{ij}^{ab} =
\upsilon_{ij}^{ab} / \Delta_{ab}^{ij}$, where
$\Delta^{ij}_{ab}=\varepsilon_i+\varepsilon_j-\varepsilon_a-\varepsilon_b$.
The correlation energy for the UEG diverges
due to the summation over
elements in the amplitudes with both the occupied orbital $i$ and the
virtual orbital $a$ close to the Fermi surface.
The rate of divergence is $\log(q)$~\cite{mattuck} and $\log[\log(q)]$~\cite{Harris}
for Hartree or HF orbital
energies, respectively. $q$ refers to the lower spherical cutoff radius in the
analytical integration over $\mathbf{q}$.
Here, we employ HF orbital energies only.
The comparison between analytic and numerical results is slightly complicated
by the fact that $\Delta^{ij}_{ab}$ approaches the analytic behavior
of $\lim_{\mathbf{q} \to 0}\Delta^{ij}_{ab}\propto |\mathbf q| \ln(|\mathbf q|)$ only slowly
with respect to the studied system size.
However, our numerical findings for $t_{ij}^{ab}$ shown in \Fig{fig:sofg}(c)
agree reasonably well with
the analytic result of $\propto |\mathbf q|^{-3}/\log(|\mathbf q|)$.
The infrared catastrophe due to the singularity of $S(\mathbf{q})$ at $|\mathbf{q}|=0$
can be inferred from the plot for MP2 theory shown in \Fig{fig:sofg}(a).

The numerical procedure follows the description in
Ref.~\cite{Shepherd2014}. We use twist averaging which
helps to reduce the fluctuations due to discretization errors of the finite
simulation cell~\cite{Gruber2018,Drummond2008,Lin2001}.

\emph{The ring summation.} ----
As already demonstrated by Macke in 1950~\cite{macke_uber_1950},
the divergence in second-order perturbation theory can be averted
by including carefully selected higher-order contributions of the many-electron
perturbation expansion, corresponding to ring diagrams.
Algebraically, this can be implemented by solving the ring-coupled-cluster amplitude equation given by
\begin{equation}
\label{eq:t-ring}
t_{ij}^{ab} = \left( \upsilon_{ij}^{ab}
             + 2  \upsilon_{ic}^{ak} t_{kj}^{cb}
             + 2  t_{ik}^{ac} \upsilon_{cj}^{kb} 
            + 4 t_{il}^{ad} \upsilon_{dc}^{lk} t_{kj}^{cb}
              \right) / \Delta_{ab}^{ij},
\end{equation}
which is formally equivalent to solving the Casida equations~\cite{scuseria_2008}.
For the ring-coupled-cluster correlation energy to converge in the long wavelength limit,
the first term on the right-hand side of Eq.(\ref{eq:t-ring}),
which is equivalent to MP2 theory, needs to be partially canceled by the additional
linear ring (lr) and quadratic ring (qr) terms in the amplitudes.

We now address the
question: How do the terms on the right-hand side in Eq.~(\ref{eq:t-ring}) cancel each other in the limit $|\mathbf{q}| \rightarrow 0$?
To this end, we rewrite Eq.~(\ref{eq:t-ring}) in the following way
\begin{equation}
\begin{split}
\label{eq:t-ring-2nd}
t^\text{rCCD}_{ij}(\mathbf{q}) = \upsilon(\mathbf{q})
  &\left(1 + 2\sum_k t^\text{rCCD}_{kj}(\mathbf{q}) + 2 \sum_k t^\text{rCCD}_{ik}(\mathbf{q}) \right. \\
          &+ \left. 4 \sum_{kl} t^\text{rCCD}_{il}(\mathbf{q}) t^\text{rCCD}_{kj}(\mathbf{q}) \right) / \Delta_{ab}^{ij} \textrm{,}
\end{split}
\end{equation}
which is possible because of momentum conservation and the fact that ring diagrams do not couple different $\mathbf{q}$.
As was shown by Freeman in Ref.~\cite{freeman_coupled-cluster_1977},
the solution of the above equation in the long wave limit leads to
$\lim_{\mathbf{q} \to 0} T^\text{rCCD}_i(\mathbf{q}) = -1/2$.
Consequently, the terms in the parenthesis in Eq.(\ref{eq:t-ring-2nd}) vanish such that
$\lim_{\mathbf{q} \to 0} t^\text{rCCD}_{ij}(\mathbf{q}) \propto {|\mathbf{q}|^{-1}}$~\cite{freeman_coupled-cluster_1977}.
We note that amplitudes diverging as $|\mathbf{q}|^{-1}$ yield convergent correlation energies
and vanishing $S(\mathbf{q})$'s for small $|\mathbf{q}|$.
The above findings are identical for Hartree and
Hartree-Fock orbital energies.
\Fig{fig:sofg}(c) depicts our numerical results for $t^\text{rCCD}_{ij}(\mathbf{q})$,
which diverges with the analytically known behavior of ${|\mathbf{q}|^{-1}}$ in the long wavelength limit.
See  Supplemental Material, which includes Refs.~\cite{Knizia2009,Valiev2010,cc4s,Gallo2021,Irmler2021,Hirata2001,kresse1996,cc4s,Kresse1999,gruneis11,hummel_2017,Irmler2021},
for numerical results confirming that $\lim_{\mathbf{q} \to 0} T^\text{rCCD}_i(\mathbf{q}) = -1/2$.

We complement the above discussion
by numerically studying the rCCD transition
structure factor and its individual diagrammatic contributions~\cite{Irmler2019}
given by
\begin{equation}
\label{eq:srccd}
S^\text{rCCD}(\mathbf{q}) = S^\text{MP2}(\mathbf{q})
       + S^\text{lr}(\mathbf{q}) + S^\text{qr}(\mathbf{q}) \text.
\end{equation}
The calculated contributions to $S(\mathbf{q})$ are depicted in
Fig.~\ref{fig:sofg}(a).  Although the individual contributions
diverge, the total rCCD transition structure factor converges towards zero in the
limit of $\mathbf{q} \rightarrow 0$ [see Fig.~\ref{fig:sofg}(a)].
We arrive at the first important insight of this work.
The singularity at $|\mathbf{q}|=0$ in $S^\text{MP2}(\mathbf{q})$ is
canceled by one-half of the linear ring terms, whereas the singularity of
the quadratic ring term is canceled by the ``other'' half of the linear term.
We also stress that the leading-order behavior of
$S^\text{rCCD}(\mathbf{q})$ in the limit $|\mathbf{q}| \rightarrow 0$
is $S^\text{rCCD}(\mathbf{q})\propto |\mathbf{q}|$. This was shown by
Bishop and L\"{u}hrmann~\cite{bishop_1982} (see Ref.~\cite{Mihm2023} for a detailed discussion).
For simulation cells with finite electron numbers this leads to a finite size error scaling
of $N^{-2/3}$, where $N$ is the number of electrons. 
Insulating systems have a different leading order behavior
around $\mathbf{q} \rightarrow 0$ given by
$S^\text{rCCD}(\mathbf{q})\propto |\mathbf{q}|^2$, implying a finite
size error scaling of $N^{-1}$~\cite{Liao2016,Martin2016}.

\emph{The coupled-cluster doubles method.} --
In addition to ring diagrams,
CCD theory includes further terms, linear and quadratic in $t^\text{CCD}_{ij}(\mathbf{q})$
(the full set of CCD equations can be found in the Supplemental Material).
This makes an analytic solution impossible even for the simple UEG model.
Note that the contributions of the linear terms are even identical to those from third-order
perturbation theory, if $t^\text{CCD}_{ij}(\mathbf{q})$ is replaced by
$\upsilon_{ij}^{ab} / \Delta_{ab}^{ij}$.
As discussed by Mattuk~\cite{mattuck} for any finite-order
perturbation theory, ring terms yield the most divergent contributions at small $|\mathbf{q}|$.
We stress that the dominance of the ring terms originates from the ``piling-up
of factors $1/q^2$,'' as the greatest piling-up occurs for the ring terms when
only a single momentum transfer is involved~\cite{gell-mann_correlation_1957}.
Therefore, this dominance prevails irrespectively of the usage of
$t^\text{CCD}_{ij}(\mathbf{q})$ instead of $\upsilon_{ij}^{ab} /
\Delta_{ab}^{ij}$ at small $|\mathbf{q}|$.
Consequently, it follows, in agreement with the work of Emrich and Zabolitzky~\cite{Emrich1984},
that for the long wavelength limit: (i) in the CCD
amplitude equations the most divergent contributions are the ones
given in Eq.(\ref{eq:t-ring-2nd}), and (ii)
these ring contributions to the CCD amplitudes must therefore cancel each other
precisely as for the rCCD amplitude equations.
This leads us to one pivotal conclusion of the
present work:
$\lim_{|\mathbf{q}| \to 0} T^\text{CCD}_i(\mathbf{q}) = \lim_{|\mathbf{q}| \to 0} T^\text{rCCD}_i(\mathbf{q}) = -1/2 $.

We corroborate the above paragraph with numerical results for the individual contributions
to $t^\text{CCD}_{ij}(\mathbf{q})$.
\Fig{fig:sofg}(c) depicts that the lr and qr contributions
diverge as $\propto {|\mathbf{q}|^{-3}/\log(q)}$. Note that these lr and qr contributions are evaluated using
CCD amplitudes.
Moreover, it is shown that the remaining linear and quadratic contributions
to the CCD amplitudes
denoted as rest-linear and rest-quadr, respectively, 
diverge with a weaker power for $\mathbf{q} \to 0$.
This underpins the conclusions drawn above that in the long wavelength limit: (i) the ring
contributions to CCD amplitudes dominate, and (ii) the ring terms cancel each other
precisely as in rCCD.
Furthermore, we find that $t^\text{CCD}_{ij}(\mathbf{q})\propto {|\mathbf{q}|^{-1}}$,
which leads to a transition structure
factor  $S^\text{CCD}(\mathbf{q})$ depicted in
\Fig{fig:sofg}(a) that approaches zero in the limit of  $\mathbf{q} \to 0$.

\emph{Triple particle-hole excitation operators} --
We now turn to CC theories that approximate the triple particle-hole
excitation operator in a perturbative manner.
In these cases the post-CCSD correlation energy and transition structure factor
contributions are given by
\begin{equation}
\label{eq:corr_triples}
        E_\text{c}^{(J)} = \sum_{\mathbf{q}}\upsilon(\mathbf{q})\underbrace{ \Biggl[\, \\
                            \frac{\delta W_{ijk}^{abc}}{\delta\upsilon( \mathbf{q} )}
                           \left( A_{abc}^{ijk} \right)_{(J)} \Biggr] }_{\coloneqq S^{(J)}(\mathbf{q})}.
\end{equation}
Here, ($J$) refers to the employed approximation.
In the case of  (T), $A_{abc}^{ijk}=\bar{W}_{abc}^{ijk}/\Delta_{abc}^{ijk}$ with
$\bar{X}^{ijk}_{abc}=\frac{4}{3}{X}^{ijk}_{abc} -2{X}^{ijk}_{acb}  +\frac{2}{3}{X}^{ijk}_{bca}  $.
$\Delta_{abc}^{ijk}$ refers to the difference in HF orbital energies for the occupied and unoccupied
states labeled by the respective indices.
$W^{abc}_{ijk}$ is defined as
\begin{equation}
\label{eq:triples_residuum}
               W_{ijk}^{abc} = P_{ijk}^{abc} \left( t_{ij}^{ae}\upsilon_{ek}^{bc} - t_{im}^{ab}\upsilon_{jk}^{mc} \right).
\end{equation}
The permutation operator $P_{ijk}^{abc}$ is defined as $P_{ijk}^{abc}X_{ijk}^{abc}= 
X_{ijk}^{abc} + X_{ikj}^{acb} + X_{kij}^{cab} + X_{kji}^{cba} + X_{jki}^{bca} + X_{jik}^{bac}$.
We employ the following functional derivative for a concise notation used to define the structure factor:
${\frac{\delta W_{ijk}^{abc}}{\delta\upsilon( \mathbf{q} )} = P_{ijk}^{abc}
                           \left( t_{ij}^{ae}\delta_{\mathbf{q},\mathbf{k}_k-\mathbf{k}_c}
                          - t_{im}^{ab}\delta_{\mathbf{q},\mathbf{k}_k-\mathbf{k}_c} \right)}$.
\Fig{fig:sofg}(b) depicts $S^\text{(T)}(\mathbf{q})$, which shares many
similarities with $S^\text{MP2}(\mathbf{q})$.
Both $S(\mathbf{q})$'s exhibit a singularity for ${|\mathbf{q}| \to 0}$
and yield correlation energies per electron that diverge as the system size increases.
The infrared catastrophe of (T) is
caused by the unscreened Coulomb interactions included in the approximation to $A_{abc}^{ijk}$~\cite{Shepherd2013}.

Here, we introduce a novel correlation energy expression
that yields convergent energies for the UEG at the level of triple particle-hole
excitation operators 
 without depending on any ad hoc parameters.
We refer to the method as (cT) because it includes the \emph{complete} set of terms 
present in the triples amplitude equations in a noniterative manner.
Naturally, the coupling of triples amplitudes with each other is disregarded.
Thus, in (cT) we use the following approximation to $A_{abc}^{ijk}=\bar{M}_{abc}^{ijk}/\Delta_{abc}^{ijk}$,
where
\begin{equation}
\label{eq:M_piecuch}
               M_{abc}^{ijk} = P_{ijk}^{abc} \left( t_{ij}^{ae}\upsilon_{ek}^{bc} + 2t_{ij}^{ae} \upsilon_{ef}^{bm} t_{mk}^{fc} + \ldots \right).
\end{equation}
For brevity we show only a selection of terms that exhibit a divergent behavior for
${\mathbf{q} \to 0}$.
Note that the (T) and (cT) approximation to $A_{abc}^{ijk}$ agrees for all terms linear in $t_{ij}^{ab}$.
To cancel the divergence in (T), (cT) includes additional ring terms 
such as the second term in the above equation. 
We show the diagrams corresponding to both terms from Eq.~(\ref{eq:M_piecuch}) in \Fig{fig:sofg}(b).
For these terms, it follows that
$\upsilon_{ek}^{bc} + 2 \upsilon_{ef}^{bm} t_{mk}^{fc}=\upsilon_{ek}^{bc}
[1+2T_m(\mathbf{q}')]$, with $\mathbf{q}' = \mathbf{k}_k - \mathbf{k}_c$.
Note that ${\mathbf{q}' \to 0}: 1+2T_m(\mathbf{q}')=0 $,
which is formally identical to our results for the cancellation of the
divergence between the linear ring and the MP2 terms in the long wavelength
limit.
We stress that (cT) also includes
additional \textit{rest} terms that are not related to the divergence of (T).
The complete expression of $M_{abc}^{ijk}$ including all contributions can be found in
the Supplemental Material and Ref.~\cite{Piecuch2002}.

As supporting numerical evidence, \Fig{fig:sofg}(b) depicts $S^\text{(cT)}(\mathbf{q})$,
which exhibits a qualitatively different behavior than $S^\text{(T)}(\mathbf{q})$ and
indicates convergence in the long wavelength limit.
Furthermore, $S^\text{(cT)-ring}(\mathbf{q})$ is shown to cancel the divergence of $S^\text{(T)}(\mathbf{q})$.
For completeness, \Fig{fig:sofg}(b) also depicts $S^\text{(cT)-rest}(\mathbf{q})$, which includes all
contributions to Eq.~(\ref{eq:M_piecuch}) that are not included in (T) or (cT)-ring,
e.g., exchangelike terms. Our findings show that $S^\text{(cT)-rest}(\mathbf{q})$ converges to zero as ${\mathbf{q} \to 0}$.
This shows that the (cT) correlation energy expression averts the infrared catastrophe of the
(T) approximation for the UEG without requiring an iterative solution of the triple amplitudes
as recently proposed in Ref.~\cite{Neufeld2023}.

\emph{Results.} ---- We now turn to the discussion of numerical results for correlation energies
obtained for electron gas simulation cells with various electron numbers at different levels of theory.
\Fig{fig:sofg}(d-e) display the behavior of the correlation energy per electron
as a function of $N^{-2/3}$ for MP2, rCCD, CCD, (T) and (cT).
The employed system sizes vary from $N=38$ electrons to $342$ electrons.
All presented correlation energies have been extrapolated to the complete basis set (CBS) limit.
The infrared catastrophe in MP2 theory becomes visible
on approach to the TDL ($N\to \infty$) for \Fig{fig:sofg}(d).
In contrast, rCCD and CCD correlation energies deviate considerably from the MP2 counterpart,
converging as $N^{-2/3}$ to the TDL.
In an analogous way, \Fig{fig:sofg}(e)
verifies that the (T) correlation energy contribution
also diverges as we move to the TDL, while its counterpart, (cT), exhibits a behavior
that more closely resembles that of rCCD and CCD theories.

Table~\ref{tab:energies} summarizes correlation energies obtained from CCD, CCD(T), and 
CCD(cT) methods compared to i-FCIQMC, DMC, and CCDT (CCD with full triple excitations) results.
The energies were extrapolated to the CBS limit, and the systems were parametrized by a range of
Wigner-Seitz radii ($r_{s}=1, 2, 3, 5, 10$ a.u.) and $N=14, 54$ electrons.
We first discuss the results obtained for the 14 electron system.
In this case i-FCIQMC can be viewed as an exact reference, whereas CCDT serves as a reference
for any approximate triples theory.
CCDT results from Ref.~\cite{Neufeld2017} are in good agreement with i-FCIQMC in the high density limit and
differ at low densities. Higher levels of CC theory are needed to capture all
important correlation effects for increasing $r_{s}$~\cite{Neufeld2017}.
CCD(T) and CCD(cT) correlation estimates are in good agreement with CCDT.
We note that the agreement between CCD(T) and CCDT is
fortuitous and only valid for small $N$,
as can be seen from the divergence of CCD(T) as a function of $N$ in
\Fig{fig:sofg}(e).

We now discuss the results obtained for the 54 electron system summarized in Table~\ref{tab:energies}.
Here, we compare to DMC reference results.
In this case CCD(T) is fortuitously close to DMC
even at relatively low densities corresponding to $r_{s}=10$~a.u.
As can be seen from \Fig{fig:sofg}(b) and \Fig{fig:sofg}(e), this agreement is due to
error cancellation between two effects. On the one hand, (T) overestimates long range correlation effects.
On the other hand, higher-order cluster operators are missing in CCD(T), which underestimates correlation
effects at lower densities.
This error cancellation fails for larger electron numbers.
CCD(cT) averts the infrared catastrophe and obtains accurate correlation energy results compared to
DMC for all densities up to $r_{s}=5$~a.u. Only at lower densities CCD(cT)
starts to deviate significantly from DMC due to the neglect of higher-order cluster operators.

 \begin{table}
\centering
       \caption{CBS limit correlation energies per electron of the UEG in mHa. r\textsubscript{s} is given in atomic units. }
       \resizebox{\columnwidth}{!}{
            \begin{tabular}{ c c c c c c c c c c}
            \hline\hline
                      & \multicolumn{4}{c}{14 electrons} &                     &  \multicolumn{4}{c}{54 electrons} \\
                        \cline{2-5}                                             \cline{7-10} 
             Method & $r\textsubscript{s}=1$ & $r\textsubscript{s}=2$ & $r\textsubscript{s}=3$ & $r\textsubscript{s}=5$  &   & $r\textsubscript{s}=1$ & $r\textsubscript{s}=2$ & $r\textsubscript{s}=5$ & $r\textsubscript{s}=10$ \\

            \hline
             CCD                 & -36.7 & -29.2 & -24.2 & -18.1    &              & -38.4 & -30.2 & -18.5 & -11.3 \\
             CCD(T)              & -37.9 & -31.5 & -27.1 & -21.4    &              & -39.9 & -33.1 & -22.6 & -15.0 \\
             CCD(cT)             & -37.8 & -31.3 & -26.9 & -21.1    &              & -39.8 & -32.8 & -22.1 & -14.5 \\
            \cline{1-10}
            CCDT\footnote{CCDT data are from the work of Neufeld and Thom~\cite{Neufeld2017}.}  & -37.9 & -31.5 & -27.0 & -21.2  \\
            i-FCIQMC\footnote{i-FCIQMC data are from the work of Shepherd \emph{et al.}~\cite{shepherd2012}}  & -38.0 & -31.8 & \textemdash & -21.9 \\
            DMC\footnote{These DMC data are from the work of L\'opez R{\'i}os \emph{et al}.~\cite{LopezRios2006}}  &&&&&& -39.0 & -32.6 & -22.8 & -15.6 \\

              \hline\hline
              \end{tabular}
     }
\label{tab:energies}
\end{table}

%
%
%

As an important test, we apply CCSD(cT) theory to a set of molecules.
In our benchmark we use a set of 26 different molecules that 
give access to 23 different closed-shell reaction energies. This selection of
molecules was already used in a previous work~\cite{Irmler2021}.
As reference we
use energies from a converged CCSDT calculation. The standard deviation of the
reaction energy for the 23 reactions is 0.9~kJ/mol for both CCSD(T) and
CCSD(cT) [the maximum error is 2.1 and 3.3~kJ/mol for CCSD(T) and
CCSD(cT), respectively]. This illustrates that both, -CCSD(T) and CCSD(cT),-
are very accurate approximations for the full CCSDT energy.

Finally, we present results for the lithium bcc metal.
We find that CCSD(cT) predicts a cohesive energy of $60.1\,$mHa/atom
in excellent agreement with the experimental estimate of $60.9\,$mHa/atom
corrected for zero-point vibrations~\cite{Schimka2011}.
Our estimate includes a HF, CCSD, (cT) and core-valence MP2 contribution
of $20.5$, $30.4$, $8$, and $1.2\,$mHa/atom, respectively.
The computational details are discussed in the Supplemental Material.
\Fig{fig:sofg}(e) depicts the (cT) correlation energy convergence.
Although our CCSD estimate of the cohesion energy is in good agreement with results from Ref.~\cite{Neufeld2023},
we note that our triples estimate is about 3 mHa/atom larger.

\emph{Summary and conclusions.} ----
We have introduced the highly accurate and computationally
efficient CCSD(cT) theory, which paves the way for achieving chemical accuracy
in \emph{ab initio} calculations of real metals.
Although the presented approach was applied to paramagnetic systems only,
it can also be generalized to ferromagnetic systems using an unrestricted formalism.
Several far reaching conclusions must be drawn from the present work.
The RPA in the electron gas
is formally identical to CCSD in the long wavelength limit.
Therefore embedding CCSD into the RPA should also be a promising approach for metals~\cite{Schaefer2021a, Schaefer2021}.
Furthermore our work could explain (part of) the discrepancies observed between DMC and CCSD(T) interaction
energies of large molecules~\cite{Al-Hamdani2021}.

The authors thankfully acknowledge support and funding from the
European Research Council (ERC) under the European Union-’s Horizon 2020 research and innovation
program (Grant Agreement No. 715594). We gratefully
acknowledge many fruitful discussions with Felix Hummel, Alejandro Gallo, and James Shepherd. 
The computational results presented have been achieved in part using the Vienna Scientific Cluster (VSC).

\bibliography{renormalized}

\end{document}


\title{
Supplemental Material for: Averting the infrared catastrophe in the gold standard of quantum chemistry
}
\author{Nikolaos Masios}
\author{Andreas Irmler}
\author{Tobias Sch\"afer}
\author{Andreas Gr\"uneis}
\email{andreas.grueneis@tuwien.ac.at}
\affiliation{
  Institute for Theoretical Physics, TU Wien,\\
  Wiedner Hauptstraße 8-10/136, 1040 Vienna, Austria
}

\maketitle

\section{Numerical results for $T_i(\mathbf{q})$}

\begin{figure*}[h]
  \centering
  \includegraphics[width=0.45\linewidth]{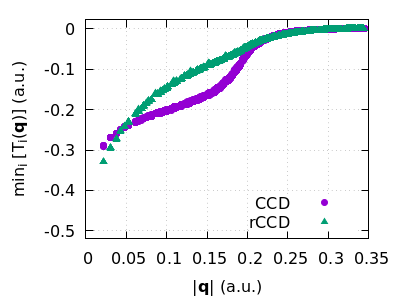}
  \caption{The minimum value for a given momentum transfer $\mathbf{q}$ is shown for the quantity $T_i(\mathbf{q})$.
           Results are shown for rCCD and CCD. Both simulations employed 730 electrons and a
           density of $r_s = 20$.}
  \label{fig:plotT}
\end{figure*}

In the main article the behaviour for the quantity $T_i(\mathbf{q})$ is
discussed for values $\mathbf{q} \to 0$. In Fig.~\ref{fig:plotT}  we show
numerical results for the UEG obtained with rCCD and CCD. One can see that
mininum value of $T_i(\mathbf{q})$ converges slowly to the theoretical expected
limit of -0.5. For rCCD results with larger electron numbers are presented
below.

\section{Structure factor for different values of $r_s$}

\begin{figure*}
  \centering
  \includegraphics[width=0.8\linewidth]{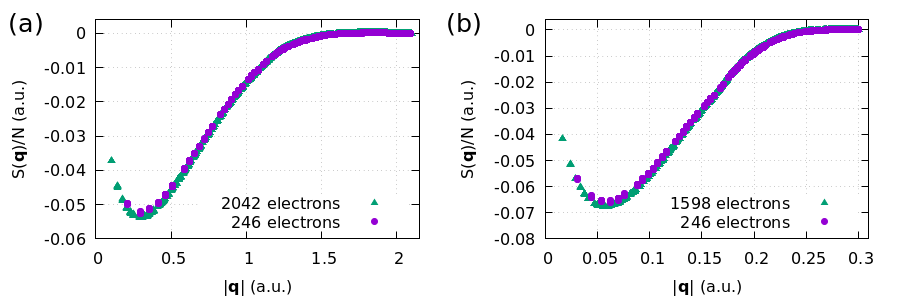}
  \caption{For two different number of electrons $N$, twist-averaged structure factors for rCCD are shown.
(a) shows results for $r_s = 3$, (b) shows results for $r_s = 20$.}
  \label{fig:rs}
\end{figure*}

The discussion in the main article was based on a structure factor obtained for a density of
$r_s = 20$. Fig.~\ref{fig:rs} shows structure factors for different values of
$r_s$. This illustrates that the discussed behaviour of the structure factor not only holds
for $r_s = 20$, but is also true for the important range of values between 1 and 5.
However, for larger values of $r_s$ the minimum of the structure factor is already observable
for smaller electron numbers. This was the reason for choosing a rather high value of $r_s$ in
the main article. In both calculations, roughly 18 virtual spatial orbitals per occupied spatial
orbital are used.

\section{rCCD results for larger electron numbers}

\begin{figure*}
  \centering
  \includegraphics[width=0.8\linewidth]{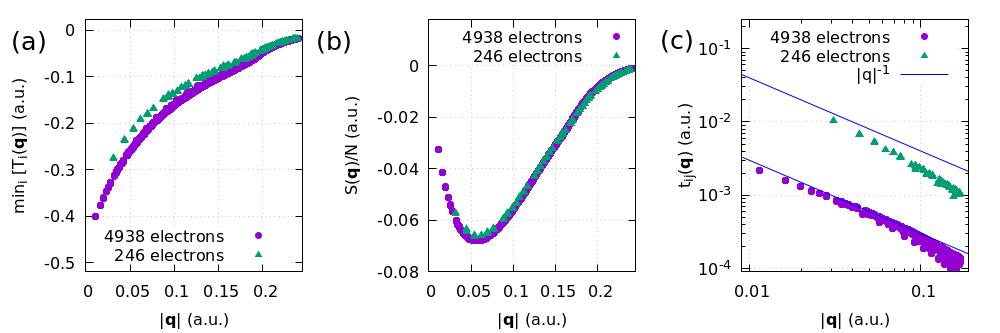}
  \caption{Results obtained using a Hartree--Fock reference for two different electron numbers $N$ and a density of $r_s = 20$.
           (a) shows $\textrm{min}_i \left[T_i(\mathbf{q})\right]$, whereas (b) shows $S(\mathbf{q})$ as
           defined in the main article. (c) shows the largest t-amplitude element with a given momentum transfer $\mathbf{q}$.  }
  \label{fig:2469occ}
\end{figure*}

Figure~\ref{fig:2469occ} shows rCCD results using $N=4938$ electrons together
with data for $N=246$ electrons.  Similar to the numerical results in the main
article, a Hartree--Fock reference is used in this calculations.
Fig.~\ref{fig:2469occ}(a) shows a very slow convergence of $\textrm{min}_i
\left[T_i(\mathbf{q}) \right] \to -0.5$ for $\mathbf{q} \to 0$, whereas (b)
shows the structure factor for both electron numbers. Finally,
Fig.~\ref{fig:2469occ}(c) shows the largest t-amplitude element on an absolute
scale for a given $\mathbf{q}$-point.  It can be seen that the theoretical
expected $1/|\mathbf{q}|$ divergence for small values of $\mathbf{q}$ sets in
only for very large electron numbers. This is a consequence of the finite size
error.  In all calculations, roughly 18 virtual spatial orbitals per occupied
spatial orbital are used.


%
%
%

\newpage
\section{Molecular testset}

Table~\ref{tab:mol} summarizes
correlation energy contributions
obtained using different triple particle-hole excitation approximations. The
geometries can be found in the work of Knizia et al. \cite{Knizia2009}. For
these calculations the Hartree--Fock ground state was obtained with the NWChem
package~\cite{Valiev2010} and interfaced to cc4s~\cite{cc4s} as described in
Ref.~\cite{Gallo2021}.  The here employed testset was already used in
a previous work \cite{Irmler2021}.  This is where Hartree--Fock and CCSD
energies for the given molecules can be found.

\begin{table}
\centering
\caption{
Triple particle-hole excitation correlation energy contributions
calculated with an aug-cc-pVTZ basis set.
(T) and (cT) are computed using a one-shot approach as
described in the main article. The energy for the full T, given in the last column, is evaluated by
$E_T = E(CCSDT) - E(CCSD)$. All energies in atomic units.}
\label{tab:mol}
\begin{tabular}{l| c| c| c}
Molecule &   (T)  &  (cT)&   T    \\
\hline
C2H2     & -0.01709& -0.01575& -0.01709 \\
C2H4     & -0.01550& -0.01440& -0.01588 \\
CH3Cl    & -0.01657& -0.01537& -0.01728 \\
CH3OH    & -0.01587& -0.01471& -0.01613 \\
CH3SH    & -0.01663& -0.01547& -0.01745 \\
CH4      & -0.00653& -0.00616& -0.00693 \\
CO       & -0.01789& -0.01642& -0.01800 \\
CO2      & -0.03031& -0.02769& -0.03001 \\
CS2      & -0.03663& -0.03330& -0.03702 \\
Cl2      & -0.01952& -0.01802& -0.02046 \\
ClF      & -0.01867& -0.01727& -0.01924 \\
F2       & -0.01966& -0.01818& -0.01967 \\
H2       &  0.00000&  0.00000&  0.00000 \\
H2O      & -0.00867& -0.00805& -0.00874 \\
H2O2     & -0.02022& -0.01861& -0.02018 \\
H2S      & -0.00867& -0.00814& -0.00941 \\
HCHO     & -0.01757& -0.01620& -0.01771 \\
HCN      & -0.01889& -0.01732& -0.01867 \\
HCOOH    & -0.02799& -0.02569& -0.02795 \\
HCl      & -0.00876& -0.00816& -0.00933 \\
HF       & -0.00755& -0.00702& -0.00758 \\
HNCO     & -0.03044& -0.02783& -0.03027 \\
N2       & -0.01981& -0.01810& -0.01944 \\
N2H4     & -0.01760& -0.01627& -0.01776 \\
NH3      & -0.00833& -0.00777& -0.00854 \\
SO2      & -0.03679& -0.03346& -0.03645 \\
\hline
\end{tabular}
\end{table}

\newpage\newpage

\section{Coupled Cluster Doubles (CCD) amplitude equation}

Here we present the closed-shell coupled-cluster doubles amplitude equations,
which can also be found elsewere\cite{Hirata2001}.
The first line contains all contributions of the rCCD method. The second line
contains all further terms which are linear in $t_{ij}^{ab}$, denoted as 
rest-linear in the main article. The other terms are further quadratic
contributions to the CCD amplitude equation, denoted as rest-quadr in the main
article.

\begin{align}
\Delta_{ij}^{ab} t_{ij}^{ab} &=
      \upsilon_{ij}^{ab}
     + P \sum_{kc} 2 \upsilon_{ic}^{ak} t_{kj}^{cb}
     + P \sum_{kcld} 2 t_{il}^{ad} v_{dc}^{lk} t_{kj}^{cb} \\
    &+ \sum_{kl} \upsilon_{ij}^{kl} t_{kl}^{ab}
     + \sum_{cd} \upsilon_{ab}^{cd} t_{ij}^{cd}
     - P \sum_{kc} \upsilon_{ci}^{ak} t_{kj}^{cb}
     - P \sum_{kc} \upsilon_{ic}^{ak} t_{kj}^{bc}
     - P \sum_{kc} \upsilon_{ci}^{bk} t_{kj}^{ac} \\
    &+ \sum_{klcd} t_{kl}^{ab} \upsilon_{cd}^{kl} t_{ij}^{cd}
     - P \sum_{ckld} 2 t_{ij}^{cb} \upsilon_{cd}^{kl} t_{kl}^{ad}
     + P \sum_{ckld} t_{ij}^{cb} \upsilon_{dc}^{kl} t_{kl}^{ad}
     - P \sum_{klcd} 2 t_{kj}^{ab} \upsilon_{cd}^{kl} t_{il}^{cd}
     + P \sum_{klcd} t_{kj}^{ab} \upsilon_{dc}^{kl} t_{il}^{cd} \\
    &+ P \sum_{kcld} \frac{1}{2} t_{il}^{da} \upsilon_{cd}^{lk} t_{kj}^{cb} 
     + P \sum_{kcld} \frac{1}{2} t_{il}^{db} \upsilon_{cd}^{lk} t_{kj}^{ac} 
     - P \sum_{kcld} t_{il}^{da} \upsilon_{dc}^{lk} t_{kj}^{cb} \\          
    &+ P \sum_{kcld} \frac{1}{2} t_{il}^{da} \upsilon_{dc}^{lk} t_{kj}^{bc} 
     - P \sum_{kcld} t_{il}^{ad} \upsilon_{dc}^{lk} t_{kj}^{bc}             
     - P \sum_{kcld} t_{il}^{ad} \upsilon_{cd}^{lk} t_{kj}^{cb}             
     + P \sum_{kcld} \frac{1}{2} t_{il}^{ad} \upsilon_{cd}^{lk} t_{kj}^{bc} 
\end{align}

\begin{equation}
        P\{\dots\}_{ij}^{ab} =  \{\dots\}_{ij}^{ab} + \{\dots\}_{ji}^{ba}
\end{equation}

\newpage
\section{Perturbative triple particle-hole excitation methods}

In this section we present the full (cT) equations complementing Eq.~(8) in the
main article.  In addition we present the equations for (T), (cT)-ring,
(cT)-rest as they are used in the article.  The following equation is used for
all employed methods
\begin{equation}
        M_{abc}^{ijk} = P_{ijk}^{abc} \left( \sum_{e} t_{ij}^{ae} J^{(X)bc}_{ek} - \sum_{m} t_{im}^{ab} J^{\text{(X)}mc}_{jk} \right),
\end{equation}
with different definitions for $J^{\text{(X)}}$ for $X \in \text{(T), (cT), (cT)-ring, (cT)-rest}$ as defined below.
We note that $J^{\text{(cT)-rest}}$ is defined implicitely via
\begin{equation}
J^{\text{(T)}} + J^{\text{(cT)-ring}} + J^{\text{(cT)-rest}} = J^{\text{(cT)}}
\end{equation}

\subsection{The (T) method}
In the (T) method the following intermediates are used
\begin{align}
J^{bc}_{ek} = \upsilon^{bc}_{ek},\\
J^{mc}_{jk} = \upsilon_{jk}^{mc}
\end{align}
\subsection{The (cT)-ring term}
For the (cT)-ring terms, the following intermediates are used
\begin{align}
J^{bc}_{ek} = \sum_{mf} 2\upsilon_{ef}^{bm} t_{km}^{cf}, \\
J^{mc}_{jk} = \sum_{nf} 2\upsilon_{jf}^{mn}t_{kn}^{cf}
\end{align}

\subsection{The (cT) method}

In the (cT) method the following intermediates are used

\begin{align}
J^{bc}_{ek} &= \upsilon^{bc}_{ek} + \sum_{f} \upsilon_{ef}^{ab}t_{k}^{f} - \sum_{m} \left( \chi^{\prime mb}_{ke}t_{m}^{c} +
              t_{m}^{b}\chi^{\prime mc}_{ek} + I_{e}^{m} t_{mk}^{bc} \right) + \sum_{mf} \left( 2\chi_{ef}^{bm} t_{km}^{cf} -
              \chi_{ef}^{bm} t_{km}^{fc} - \chi_{fe}^{bm}t_{mk}^{fc} - t_{km}^{fb}\chi_{fe}^{cm} \right) + \sum_{mn} t_{mn}^{cb}\chi_{ke}^{mn}, \\
J^{mc}_{jk} &= \upsilon_{jk}^{mc} - \sum_{n}\upsilon_{jk}^{mn}t_{n}^{c} + \sum_{f} \left( \chi^{\prime\prime mc}_{jf}t^{f}_{k} +
              t_{j}^{f}\chi^{\prime\prime mc}_{fk} \right) + \sum_{nf} \left( 2\chi_{jf}^{mn}t_{kn}^{cf} - \chi_{jf}^{mn}t_{kn}^{fc}-
              \chi_{jf}^{nm}t_{nk}^{fc} - t_{nj}^{cf}\chi_{kf}^{nm} \right) + \sum_{ef} t_{kj}^{ef}\chi_{ef}^{cm}, \\
\chi^{\prime mb}_{ke} &= \upsilon_{ke}^{mb} - \frac{1}{2}\sum_{n}\upsilon^{mn}_{ke}t_{n}^{b} + \sum_{f}t_{k}^{f}\chi^{\prime bm}_{ef}, \\
\chi^{\prime bm}_{ef} &= \upsilon^{bm}_{ef} - \frac{1}{2}\sum_{n} t^{b}_{n}\upsilon^{nm}_{ef}, \\
\chi^{\prime mc}_{ek} &= \upsilon_{ek}^{mc} - \frac{1}{2}\sum_{n}\upsilon^{mn}_{ek}t_{n}^{c} + \sum_{f}\chi^{\prime cm}_{fe}t_{k}^{f}, \\
I^{m}_{c} &= f_{c}^{m} + \sum_{nf} \left( 2\upsilon_{cf}^{mn}t_{n}^{f} - \upsilon_{fc}^{mn}t_{n}^{f}\right), \\
\chi_{ef}^{bm} &= \chi^{\prime bm}_{ef} - \frac{1}{2}\sum_{n}t_{n}^{b}\upsilon^{nm}_{ef}, \\
\chi_{ke}^{mn} &= \upsilon^{mn}_{ke} + \sum_{f}t_{k}^{f}\upsilon^{mn}_{fe}, \\
\chi^{\prime\prime mc}_{jf} &= \upsilon^{mc}_{jf} - \sum_{n}\upsilon^{mn}_{jf}t_{n}^{c} +\frac{1}{2}\sum_{g}t_{j}^{g}\chi^{cm}_{fg}, \\
\chi^{\prime\prime mc}_{fk} &= \upsilon^{mc}_{fk} - \sum_{n}\upsilon^{mn}_{fk}t_{n}^{c} +\frac{1}{2}\sum_{g}\chi^{cm}_{gf}t^{g}_{k},
\end{align}

\newpage

\section{Computational details on the calculations of metallic lithium}

Metallic body-centered cubic (BCC) lithium is considered.
All calculations are performed using the Vienna Ab initio Simulation Package (VASP)~\cite{kresse1996} for HF and MP2 energies, as well as cc4s code~\cite{cc4s} for coupled cluster energies. 
VASP is based on periodic boundary conditions and makes use of the projector augmented-wave (PAW) formalism~\cite{Kresse1999}.
We consider one valence electron per Li atom, using the PAW POTCAR file labeled as $\texttt{Li\_GW}$ and a plane-wave cutoff parameter of $\texttt{ENCUT} = 141\,\text{eV}$.

For efficient coupled cluster calculations, a compression of the unoccupied HF manifold is achieved via approximate MP2 natural orbitals (NOs)~\cite{gruneis11}, while a low-rank decomposition is employed to compress the Coulomb integrals as described in Ref.~\cite{hummel_2017}.
Both schemes introduce a controlable error which is kept well below $10^{-2}$ mHa for the low-rank decomposition of the Coulomb integrals.
The basis-set error of the CCSD correlation contribution, as introduced by the restricted number of NOs, is converged to sub-mHa accuracy using a recently proposed focal-point correction scheme (here denoted as FPC)~\cite{Irmler2021}, as can be observed in Fig. \ref{fig:Li}(b).
For the (cT) and (T) correlation contribution, a relatively small number of 5 NOs per occupied orbital is used and corrected by an $1/N_{NO}$ extrapolation from 5 to 10 NOs using a simulation cell containing 26 Li atoms.
This is sufficient, since the basis-set dependence of both (cT) and (T) is weak.

Furthermore, the core-valence contribution to the correlation energy is estimated by an all-electron PAW POTCAR file labeled as $\texttt{Li\_AE\_GW}$. In this case three electrons per Li atom are considered as valence electrons. Since the $1\text s^2$ HF orbitals are located about $-63\,\text{eV}$ below the Fermi level, we consider the MP2 method as sufficient for this task. Here a plane-wave cutoff of $\texttt{ENCUT} = 543\,\text{eV}$ is used. The individual contributions to the atomization energy of bulk bcc lithium are reported in the main document. 

\section{Thermodynamic limit extrapolation}

Fig \ref{fig:Li}(a)+(c) depicts the thermodynamic limit extrapolation of the HF and CCSD correlation energies of  bulk bcc lithium.
The corresponding numerical data is summarized in Tab.\ref{tab:dat1}. 
We find that using the extrapolation law of $(N_kN_\text{atoms})^{-2/3}$ fits the computed energies and enables a precise extrapolation.
Here $N_k$ denotes the number of k-points used to sample the first Brillouin zone. We note that Brillouin zone sampling
with more than one k-point is not yet supported in cc4s.
Therefore all post-HF calculations employ $N_k=1$ and large supercells with geometries given below.
However, twist-averaging is used to reduce the fluctuations due to discretization errors of the finite simulation cells. The geometries and
atomic structures of these cells are given below.
The employed power law for the extrapolation to the limit $N\rightarrow\infty$ is well justified by the linear behaviour of the
CCSD structure factor in the long wave length limit $\mathbf{q}\rightarrow0$. However, this also implies that the system sizes included
in such an extrapolation have to be large enough to sample sufficiently short wavevectors in reciprocal space where the
structure factor approaches $\mathbf{q}\rightarrow0$ linearly.

The extrapolation of the (cT) correlation energy contribution for lithium shown in the main article, employs an $(N_\text{atoms})^{-2/3}$ fit and system sizes of 128 and 208 atoms.
We note, however, that the (cT) correlation energies obtained for smaller systems deviate from the employed extrapolation law. To estimate the
remaining uncertainty in the thermodynamic limit extrapolation of the (cT) energy we have performed the following analysis based on the UEG.

\begin{figure}[H]
  \centering
  \includegraphics[width=0.6\linewidth]{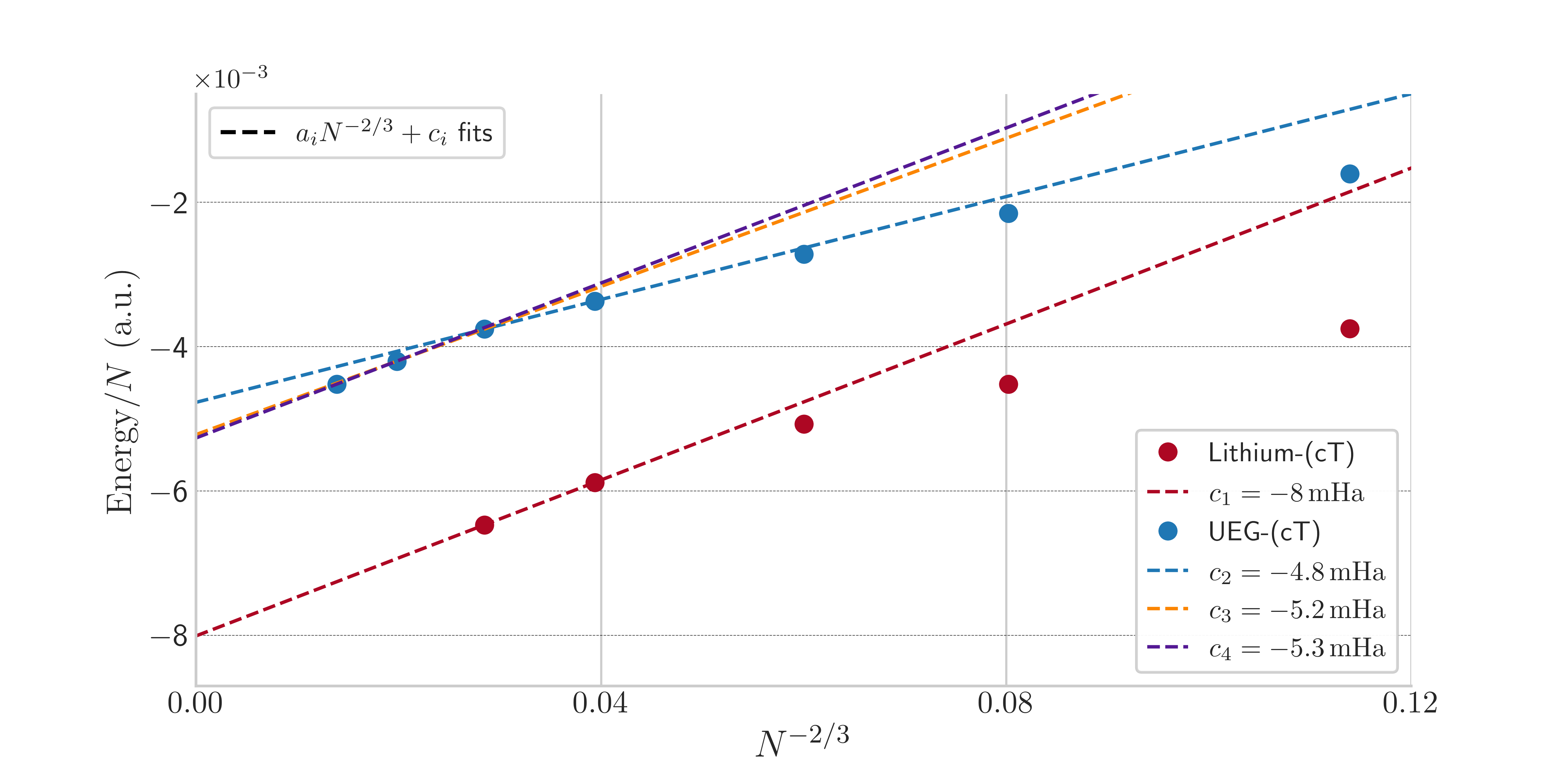}
  \caption{Thermodynamic limit convergence of the (cT) energy per electron.}
  \label{fig:UEG_TDL}
\end{figure}

\begin{figure}[H]
  \centering
  \includegraphics[width=1.0\linewidth]{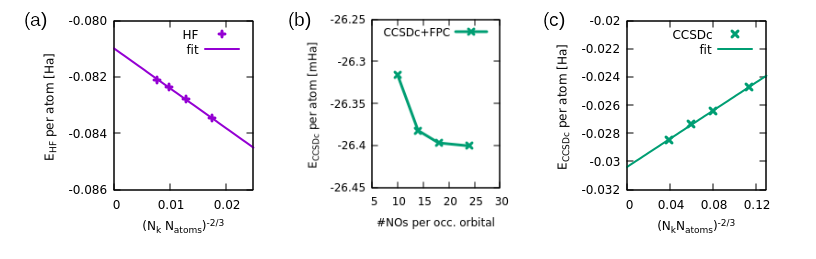}
  \caption{(a) Thermodynamic limit convergence of the HF energy per atom. (b) Basis-set convergence of the CCSD correlation energy per atom using a simulation cell with 44 Li atoms. (c) Thermodynamic limit convergence of the CCSD correlation energy per atom.}
  \label{fig:Li}
\end{figure}

Fig.~\ref{fig:UEG_TDL} depicts the (cT) correlation energies of the UEG for a range of electron numbers at a density corresponding to $r_{\rm s}$=3.2. This density and the
employed basis set sizes are identical to the calculations for lithium.
Our findings indicate that the difference between thermodynamic limit extrapolations employing the $(N)^{-2/3}$ fits and system sizes corresponding
to 358-610 electrons and 128-208 electrons deviate by less than 0.5~mHa/electron.
Furthermore the extrapolated estimates using 358-610 electrons and 208-358 electrons correspond to -5.2~mHa/electron and -5.3~mHa/electron, deviating by less than 0.1~mHa/electron.
From this we estimate that the remaining uncertainty in our estimate of the (cT) contribution to the 
lithium cohesive energy obtained from extrapolations using 128 and 208 atoms is below 1~mHa/atom.
Tab.~\ref{tab:dat2} summarizes the numerical data of (cT) energies for the UEG and lithium.

\begin{table}
    \centering
    \begin{tabular}{c|c}
         $N_kN_\text{atoms}$& HF  \\
	 \hline
         216 & -0.08344  \\
         343 & -0.08280  \\
         512 & -0.08237  \\
         729 & -0.08210  \\
    \end{tabular}
    \qquad
    \begin{tabular}{c|c}
         $N_kN_\text{atoms}$& CCSDc  \\
	 \hline
          26 & -0.02463  \\
          44 & -0.02633  \\
          68 & -0.02727  \\
         128 & -0.02841  \\
    \end{tabular}
    \caption{Total energies of the plots (a) and (c) of Fig. \ref{fig:Li} of this Supplemental Material. The energies are in Ha per atom.}
    \label{tab:dat1}
\end{table}

\begin{table}
    \centering
    \begin{tabular}{c|c|c}
         $N_\text{elect}$& UEG-(cT) & Li-(cT)  \\
	 \hline
          26 & -1.60 & -3.75 \\
          44 & -2.15 & -4.52 \\
          68 & -2.72 & -5.07 \\
         128 & -3.37 & -5.88 \\
         208 & -3.76 & -6.47 \\
         358 & -4.20 &  ---  \\
         610 & -4.53 &  ---  \\
    \end{tabular}
    \caption{Tabulated correlation contribution at the level of (cT) to the atomization energy of Li as depicted in Fig. 1 of the main manuscript.
 UEG-(cT) energies are discussed in the text. All energies in mHa per electron.}
    \label{tab:dat2}
\end{table}

\subsection*{Atomic structures of the used bcc lithium supercells in units of {\AA}ngstrom.}

\subsubsection*{26 atoms}
\begin{footnotesize}
\begin{verbatim}
  volume of cell :     535.2100

  lattice vectors of the cell                    
     6.905952995 -3.452976497  3.452976497  
     0.000000000  6.905952995  3.452976497  
    -3.452976497 -3.452976497  6.905952995  

  length of vectors
     8.458030512  7.721090173  8.458030512    

  position of ions in fractional coordinates 
     0.000000000  0.000000000  0.000000000
     0.923076923  0.384615385  0.846153846
     0.615384615  0.923076923  0.230769231
     0.461538462  0.692307692  0.923076923
     0.846153846  0.769230769  0.692307692
     0.153846154  0.230769231  0.307692308
     0.538461538  0.307692308  0.076923077
     0.230769231  0.846153846  0.461538462
     0.384615385  0.076923077  0.769230769
     0.769230769  0.153846154  0.538461538
     0.076923077  0.615384615  0.153846154
     0.307692308  0.461538462  0.615384615
     0.692307692  0.538461538  0.384615385
     0.230769231  0.346153846  0.961538462
     0.153846154  0.730769231  0.807692308
     0.846153846  0.269230769  0.192307692
     0.692307692  0.038461538  0.884615385
     0.076923077  0.115384615  0.653846154
     0.384615385  0.576923077  0.269230769
     0.769230769  0.653846154  0.038461538
     0.461538462  0.192307692  0.423076923
     0.615384615  0.423076923  0.730769231
     0.000000000  0.500000000  0.500000000
     0.307692308  0.961538462  0.115384615
     0.538461538  0.807692308  0.576923077
     0.923076923  0.884615385  0.346153846
\end{verbatim}
\end{footnotesize}

\subsubsection*{44 atoms}
\begin{footnotesize}
\begin{verbatim}
  volume of cell :     905.7400

  lattice vectors of the cell
     3.452976497 -6.905952995  6.905952995
     0.000000000  6.905952995  6.905952995
    -6.905952995 -3.452976497  6.905952995

  length of vectors
    10.358929492  9.766492386 10.358929492

  position of ions in fractional coordinates 
     0.000000000  0.000000000  0.000000000
     0.454545455  0.818181818  0.727272727
     0.727272727  0.409090909  0.363636364
     0.727272727  0.909090909  0.363636364
     0.181818182  0.227272727  0.090909091
     0.909090909  0.636363636  0.454545455
     0.090909091  0.863636364  0.545454545
     0.363636364  0.454545455  0.181818182
     0.818181818  0.272727273  0.909090909
     0.272727273  0.090909091  0.636363636
     0.545454545  0.681818182  0.272727273
     0.363636364  0.954545455  0.181818182
     0.000000000  0.500000000  0.000000000
     0.454545455  0.318181818  0.727272727
     0.181818182  0.727272727  0.090909091
     0.636363636  0.545454545  0.818181818
     0.909090909  0.136363636  0.454545455
     0.090909091  0.363636364  0.545454545
     0.818181818  0.772727273  0.909090909
     0.272727273  0.590909091  0.636363636
     0.545454545  0.181818182  0.272727273
     0.636363636  0.045454545  0.818181818
     0.136363636  0.295454545  0.818181818
     0.590909091  0.113636364  0.545454545
     0.863636364  0.704545455  0.181818182
     0.863636364  0.204545455  0.181818182
     0.318181818  0.522727273  0.909090909
     0.045454545  0.931818182  0.272727273
     0.227272727  0.159090909  0.363636364
     0.500000000  0.750000000  0.000000000
     0.954545455  0.568181818  0.727272727
     0.409090909  0.386363636  0.454545455
     0.681818182  0.977272727  0.090909091
     0.500000000  0.250000000  0.000000000
     0.136363636  0.795454545  0.818181818
     0.590909091  0.613636364  0.545454545
     0.318181818  0.022727273  0.909090909
     0.772727273  0.840909091  0.636363636
     0.045454545  0.431818182  0.272727273
     0.227272727  0.659090909  0.363636364
     0.954545455  0.068181818  0.727272727
     0.409090909  0.886363636  0.454545455
     0.681818182  0.477272727  0.090909091
     0.772727273  0.340909091  0.636363636
\end{verbatim}
\end{footnotesize}

\subsubsection*{68 atoms}
\begin{footnotesize}
\begin{verbatim}
  volume of cell :    1399.7800

  lattice vectors of the cell                    
    10.358929492  3.452976497  3.452976497  
     0.000000000  6.905952995 10.358929492  
     3.452976497-10.358929492  3.452976497  

  length of vectors
    11.452227451 12.449883814 11.452227451  

  position of ions in fractional coordinates 
     0.000000000  0.000000000  0.000000000
     0.176470588  0.117647059  0.470588235
     0.352941176  0.235294118  0.941176471
     0.029411765  0.352941176  0.911764706
     0.852941176  0.235294118  0.441176471
     0.529411765  0.352941176  0.411764706
     0.705882353  0.470588235  0.882352941
     0.205882353  0.470588235  0.382352941
     0.382352941  0.588235294  0.852941176
     0.058823529  0.705882353  0.823529412
     0.882352941  0.588235294  0.352941176
     0.558823529  0.705882353  0.323529412
     0.735294118  0.823529412  0.794117647
     0.235294118  0.823529412  0.294117647
     0.411764706  0.941176471  0.764705882
     0.088235294  0.058823529  0.735294118
     0.911764706  0.941176471  0.264705882
     0.588235294  0.058823529  0.235294118
     0.764705882  0.176470588  0.705882353
     0.264705882  0.176470588  0.205882353
     0.441176471  0.294117647  0.676470588
     0.117647059  0.411764706  0.647058824
     0.941176471  0.294117647  0.176470588
     0.617647059  0.411764706  0.147058824
     0.794117647  0.529411765  0.617647059
     0.294117647  0.529411765  0.117647059
     0.470588235  0.647058824  0.588235294
     0.147058824  0.764705882  0.558823529
     0.970588235  0.647058824  0.088235294
     0.647058824  0.764705882  0.058823529
     0.823529412  0.882352941  0.529411765
     0.323529412  0.882352941  0.029411765
     0.500000000  0.000000000  0.500000000
     0.676470588  0.117647059  0.970588235
     0.176470588  0.117647059  0.970588235
     0.352941176  0.235294118  0.441176471
     0.529411765  0.352941176  0.911764706
     0.205882353  0.470588235  0.882352941
     0.029411765  0.352941176  0.411764706
     0.705882353  0.470588235  0.382352941
     0.882352941  0.588235294  0.852941176
     0.382352941  0.588235294  0.352941176
     0.558823529  0.705882353  0.823529412
     0.235294118  0.823529412  0.794117647
     0.058823529  0.705882353  0.323529412
     0.735294118  0.823529412  0.294117647
     0.911764706  0.941176471  0.764705882
     0.411764706  0.941176471  0.264705882
     0.588235294  0.058823529  0.735294118
     0.264705882  0.176470588  0.705882353
     0.088235294  0.058823529  0.235294118
     0.764705882  0.176470588  0.205882353
     0.941176471  0.294117647  0.676470588
     0.441176471  0.294117647  0.176470588
     0.617647059  0.411764706  0.647058824
     0.294117647  0.529411765  0.617647059
     0.117647059  0.411764706  0.147058824
     0.794117647  0.529411765  0.117647059
     0.970588235  0.647058824  0.588235294
     0.470588235  0.647058824  0.088235294
     0.647058824  0.764705882  0.558823529
     0.323529412  0.882352941  0.529411765
     0.147058824  0.764705882  0.058823529
     0.823529412  0.882352941  0.029411765
     0.000000000  0.000000000  0.500000000
     0.500000000  0.000000000  0.000000000
     0.676470588  0.117647059  0.470588235
     0.852941176  0.235294118  0.941176471
\end{verbatim}
\end{footnotesize}

\subsubsection*{128 atoms}
\begin{footnotesize}
\begin{verbatim}
  volume of cell :    2634.8800

  lattice vectors of the cell                    
    13.811905989  0.000000000  0.000000000  
     0.000000000 13.811905989  0.000000000  
     0.000000000  0.000000000 13.811905989  

  length of vectors
    13.811905989 13.811905989 13.811905989  

  position of ions in fractional coordinates 
     0.000000000  0.000000000  0.000000000
     0.000000000  0.000000000  0.250000000
     0.000000000  0.000000000  0.500000000
     0.000000000  0.000000000  0.750000000
     0.000000000  0.250000000  0.000000000
     0.000000000  0.250000000  0.250000000
     0.000000000  0.250000000  0.500000000
     0.000000000  0.250000000  0.750000000
     0.000000000  0.500000000  0.000000000
     0.000000000  0.500000000  0.250000000
     0.000000000  0.500000000  0.500000000
     0.000000000  0.500000000  0.750000000
     0.000000000  0.750000000  0.000000000
     0.000000000  0.750000000  0.250000000
     0.000000000  0.750000000  0.500000000
     0.000000000  0.750000000  0.750000000
     0.250000000  0.000000000  0.000000000
     0.250000000  0.000000000  0.250000000
     0.250000000  0.000000000  0.500000000
     0.250000000  0.000000000  0.750000000
     0.250000000  0.250000000  0.000000000
     0.250000000  0.250000000  0.250000000
     0.250000000  0.250000000  0.500000000
     0.250000000  0.250000000  0.750000000
     0.250000000  0.500000000  0.000000000
     0.250000000  0.500000000  0.250000000
     0.250000000  0.500000000  0.500000000
     0.250000000  0.500000000  0.750000000
     0.250000000  0.750000000  0.000000000
     0.250000000  0.750000000  0.250000000
     0.250000000  0.750000000  0.500000000
     0.250000000  0.750000000  0.750000000
     0.500000000  0.000000000  0.000000000
     0.500000000  0.000000000  0.250000000
     0.500000000  0.000000000  0.500000000
     0.500000000  0.000000000  0.750000000
     0.500000000  0.250000000  0.000000000
     0.500000000  0.250000000  0.250000000
     0.500000000  0.250000000  0.500000000
     0.500000000  0.250000000  0.750000000
     0.500000000  0.500000000  0.000000000
     0.500000000  0.500000000  0.250000000
     0.500000000  0.500000000  0.500000000
     0.500000000  0.500000000  0.750000000
     0.500000000  0.750000000  0.000000000
     0.500000000  0.750000000  0.250000000
     0.500000000  0.750000000  0.500000000
     0.500000000  0.750000000  0.750000000
     0.750000000  0.000000000  0.000000000
     0.750000000  0.000000000  0.250000000
     0.750000000  0.000000000  0.500000000
     0.750000000  0.000000000  0.750000000
     0.750000000  0.250000000  0.000000000
     0.750000000  0.250000000  0.250000000
     0.750000000  0.250000000  0.500000000
     0.750000000  0.250000000  0.750000000
     0.750000000  0.500000000  0.000000000
     0.750000000  0.500000000  0.250000000
     0.750000000  0.500000000  0.500000000
     0.750000000  0.500000000  0.750000000
     0.750000000  0.750000000  0.000000000
     0.750000000  0.750000000  0.250000000
     0.750000000  0.750000000  0.500000000
     0.750000000  0.750000000  0.750000000
     0.125000000  0.125000000  0.125000000
     0.125000000  0.125000000  0.375000000
     0.125000000  0.125000000  0.625000000
     0.125000000  0.125000000  0.875000000
     0.125000000  0.375000000  0.125000000
     0.125000000  0.375000000  0.375000000
     0.125000000  0.375000000  0.625000000
     0.125000000  0.375000000  0.875000000
     0.125000000  0.625000000  0.125000000
     0.125000000  0.625000000  0.375000000
     0.125000000  0.625000000  0.625000000
     0.125000000  0.625000000  0.875000000
     0.125000000  0.875000000  0.125000000
     0.125000000  0.875000000  0.375000000
     0.125000000  0.875000000  0.625000000
     0.125000000  0.875000000  0.875000000
     0.375000000  0.125000000  0.125000000
     0.375000000  0.125000000  0.375000000
     0.375000000  0.125000000  0.625000000
     0.375000000  0.125000000  0.875000000
     0.375000000  0.375000000  0.125000000
     0.375000000  0.375000000  0.375000000
     0.375000000  0.375000000  0.625000000
     0.375000000  0.375000000  0.875000000
     0.375000000  0.625000000  0.125000000
     0.375000000  0.625000000  0.375000000
     0.375000000  0.625000000  0.625000000
     0.375000000  0.625000000  0.875000000
     0.375000000  0.875000000  0.125000000
     0.375000000  0.875000000  0.375000000
     0.375000000  0.875000000  0.625000000
     0.375000000  0.875000000  0.875000000
     0.625000000  0.125000000  0.125000000
     0.625000000  0.125000000  0.375000000
     0.625000000  0.125000000  0.625000000
     0.625000000  0.125000000  0.875000000
     0.625000000  0.375000000  0.125000000
     0.625000000  0.375000000  0.375000000
     0.625000000  0.375000000  0.625000000
     0.625000000  0.375000000  0.875000000
     0.625000000  0.625000000  0.125000000
     0.625000000  0.625000000  0.375000000
     0.625000000  0.625000000  0.625000000
     0.625000000  0.625000000  0.875000000
     0.625000000  0.875000000  0.125000000
     0.625000000  0.875000000  0.375000000
     0.625000000  0.875000000  0.625000000
     0.625000000  0.875000000  0.875000000
     0.875000000  0.125000000  0.125000000
     0.875000000  0.125000000  0.375000000
     0.875000000  0.125000000  0.625000000
     0.875000000  0.125000000  0.875000000
     0.875000000  0.375000000  0.125000000
     0.875000000  0.375000000  0.375000000
     0.875000000  0.375000000  0.625000000
     0.875000000  0.375000000  0.875000000
     0.875000000  0.625000000  0.125000000
     0.875000000  0.625000000  0.375000000
     0.875000000  0.625000000  0.625000000
     0.875000000  0.625000000  0.875000000
     0.875000000  0.875000000  0.125000000
     0.875000000  0.875000000  0.375000000
     0.875000000  0.875000000  0.625000000
     0.875000000  0.875000000  0.875000000
\end{verbatim}
\end{footnotesize}

\subsubsection*{208 atoms}
\begin{footnotesize}
\begin{verbatim}
  volume of cell :    4281.6800

  lattice vectors of the cell                    
    13.811905989 -6.905952995  6.905952995  
     0.000000000 13.811905989  6.905952995  
    -6.905952995 -6.905952995 13.811905989  

  length of vectors
    16.916061024 15.442180345 16.916061024  

  position of ions in fractional coordinates 
     0.000000000  0.000000000  0.000000000
     0.923076923  0.384615385  0.846153846
     0.615384615  0.923076923  0.230769231
     0.461538462  0.692307692  0.923076923
     0.846153846  0.769230769  0.692307692
     0.153846154  0.230769231  0.307692308
     0.538461538  0.307692308  0.076923077
     0.230769231  0.846153846  0.461538462
     0.384615385  0.076923077  0.769230769
     0.769230769  0.153846154  0.538461538
     0.076923077  0.615384615  0.153846154
     0.307692308  0.461538462  0.615384615
     0.692307692  0.538461538  0.384615385
     0.076923077  0.115384615  0.153846154
     0.000000000  0.500000000  0.000000000
     0.692307692  0.038461538  0.384615385
     0.538461538  0.807692308  0.076923077
     0.923076923  0.884615385  0.846153846
     0.230769231  0.346153846  0.461538462
     0.615384615  0.423076923  0.230769231
     0.307692308  0.961538462  0.615384615
     0.461538462  0.192307692  0.923076923
     0.846153846  0.269230769  0.692307692
     0.153846154  0.730769231  0.307692308
     0.384615385  0.576923077  0.769230769
     0.769230769  0.653846154  0.538461538
     0.961538462  0.192307692  0.923076923
     0.884615385  0.576923077  0.769230769
     0.576923077  0.115384615  0.153846154
     0.423076923  0.884615385  0.846153846
     0.807692308  0.961538462  0.615384615
     0.115384615  0.423076923  0.230769231
     0.500000000  0.500000000  0.000000000
     0.192307692  0.038461538  0.384615385
     0.346153846  0.269230769  0.692307692
     0.730769231  0.346153846  0.461538462
     0.038461538  0.807692308  0.076923077
     0.269230769  0.653846154  0.538461538
     0.653846154  0.730769231  0.307692308
     0.038461538  0.307692308  0.076923077
     0.961538462  0.692307692  0.923076923
     0.653846154  0.230769231  0.307692308
     0.500000000  0.000000000  0.000000000
     0.884615385  0.076923077  0.769230769
     0.192307692  0.538461538  0.384615385
     0.576923077  0.615384615  0.153846154
     0.269230769  0.153846154  0.538461538
     0.423076923  0.384615385  0.846153846
     0.807692308  0.461538462  0.615384615
     0.115384615  0.923076923  0.230769231
     0.346153846  0.769230769  0.692307692
     0.730769231  0.846153846  0.461538462
     0.192307692  0.038461538  0.884615385
     0.115384615  0.423076923  0.730769231
     0.807692308  0.961538462  0.115384615
     0.653846154  0.730769231  0.807692308
     0.038461538  0.807692308  0.576923077
     0.346153846  0.269230769  0.192307692
     0.730769231  0.346153846  0.961538462
     0.423076923  0.884615385  0.346153846
     0.576923077  0.115384615  0.653846154
     0.961538462  0.192307692  0.423076923
     0.269230769  0.653846154  0.038461538
     0.500000000  0.500000000  0.500000000
     0.884615385  0.576923077  0.269230769
     0.269230769  0.153846154  0.038461538
     0.192307692  0.538461538  0.884615385
     0.884615385  0.076923077  0.269230769
     0.730769231  0.846153846  0.961538462
     0.115384615  0.923076923  0.730769231
     0.423076923  0.384615385  0.346153846
     0.807692308  0.461538462  0.115384615
     0.500000000  0.000000000  0.500000000
     0.653846154  0.230769231  0.807692308
     0.038461538  0.307692308  0.576923077
     0.346153846  0.769230769  0.192307692
     0.576923077  0.615384615  0.653846154
     0.961538462  0.692307692  0.423076923
     0.153846154  0.230769231  0.807692308
     0.076923077  0.615384615  0.653846154
     0.769230769  0.153846154  0.038461538
     0.615384615  0.923076923  0.730769231
     0.000000000  0.000000000  0.500000000
     0.307692308  0.461538462  0.115384615
     0.692307692  0.538461538  0.884615385
     0.384615385  0.076923077  0.269230769
     0.538461538  0.307692308  0.576923077
     0.923076923  0.384615385  0.346153846
     0.230769231  0.846153846  0.961538462
     0.461538462  0.692307692  0.423076923
     0.846153846  0.769230769  0.192307692
     0.230769231  0.346153846  0.961538462
     0.153846154  0.730769231  0.807692308
     0.846153846  0.269230769  0.192307692
     0.692307692  0.038461538  0.884615385
     0.076923077  0.115384615  0.653846154
     0.384615385  0.576923077  0.269230769
     0.769230769  0.653846154  0.038461538
     0.461538462  0.192307692  0.423076923
     0.615384615  0.423076923  0.730769231
     0.000000000  0.500000000  0.500000000
     0.307692308  0.961538462  0.115384615
     0.538461538  0.807692308  0.576923077
     0.923076923  0.884615385  0.346153846
     0.115384615  0.173076923  0.980769231
     0.038461538  0.557692308  0.826923077
     0.730769231  0.096153846  0.211538462
     0.576923077  0.865384615  0.903846154
     0.961538462  0.942307692  0.673076923
     0.269230769  0.403846154  0.288461538
     0.653846154  0.480769231  0.057692308
     0.346153846  0.019230769  0.442307692
     0.500000000  0.250000000  0.750000000
     0.884615385  0.326923077  0.519230769
     0.192307692  0.788461538  0.134615385
     0.423076923  0.634615385  0.596153846
     0.807692308  0.711538462  0.365384615
     0.192307692  0.288461538  0.134615385
     0.115384615  0.673076923  0.980769231
     0.807692308  0.211538462  0.365384615
     0.653846154  0.980769231  0.057692308
     0.038461538  0.057692308  0.826923077
     0.346153846  0.519230769  0.442307692
     0.730769231  0.596153846  0.211538462
     0.423076923  0.134615385  0.596153846
     0.576923077  0.365384615  0.903846154
     0.961538462  0.442307692  0.673076923
     0.269230769  0.903846154  0.288461538
     0.500000000  0.750000000  0.750000000
     0.884615385  0.826923077  0.519230769
     0.076923077  0.365384615  0.903846154
     0.000000000  0.750000000  0.750000000
     0.692307692  0.288461538  0.134615385
     0.538461538  0.057692308  0.826923077
     0.923076923  0.134615385  0.596153846
     0.230769231  0.596153846  0.211538462
     0.615384615  0.673076923  0.980769231
     0.307692308  0.211538462  0.365384615
     0.461538462  0.442307692  0.673076923
     0.846153846  0.519230769  0.442307692
     0.153846154  0.980769231  0.057692308
     0.384615385  0.826923077  0.519230769
     0.769230769  0.903846154  0.288461538
     0.153846154  0.480769231  0.057692308
     0.076923077  0.865384615  0.903846154
     0.769230769  0.403846154  0.288461538
     0.615384615  0.173076923  0.980769231
     0.000000000  0.250000000  0.750000000
     0.307692308  0.711538462  0.365384615
     0.692307692  0.788461538  0.134615385
     0.384615385  0.326923077  0.519230769
     0.538461538  0.557692308  0.826923077
     0.923076923  0.634615385  0.596153846
     0.230769231  0.096153846  0.211538462
     0.461538462  0.942307692  0.673076923
     0.846153846  0.019230769  0.442307692
     0.307692308  0.211538462  0.865384615
     0.230769231  0.596153846  0.711538462
     0.923076923  0.134615385  0.096153846
     0.769230769  0.903846154  0.788461538
     0.153846154  0.980769231  0.557692308
     0.461538462  0.442307692  0.173076923
     0.846153846  0.519230769  0.942307692
     0.538461538  0.057692308  0.326923077
     0.692307692  0.288461538  0.634615385
     0.076923077  0.365384615  0.403846154
     0.384615385  0.826923077  0.019230769
     0.615384615  0.673076923  0.480769231
     0.000000000  0.750000000  0.250000000
     0.384615385  0.326923077  0.019230769
     0.307692308  0.711538462  0.865384615
     0.000000000  0.250000000  0.250000000
     0.846153846  0.019230769  0.942307692
     0.230769231  0.096153846  0.711538462
     0.538461538  0.557692308  0.326923077
     0.923076923  0.634615385  0.096153846
     0.615384615  0.173076923  0.480769231
     0.769230769  0.403846154  0.788461538
     0.153846154  0.480769231  0.557692308
     0.461538462  0.942307692  0.173076923
     0.692307692  0.788461538  0.634615385
     0.076923077  0.865384615  0.403846154
     0.269230769  0.403846154  0.788461538
     0.192307692  0.788461538  0.634615385
     0.884615385  0.326923077  0.019230769
     0.730769231  0.096153846  0.711538462
     0.115384615  0.173076923  0.480769231
     0.423076923  0.634615385  0.096153846
     0.807692308  0.711538462  0.865384615
     0.500000000  0.250000000  0.250000000
     0.653846154  0.480769231  0.557692308
     0.038461538  0.557692308  0.326923077
     0.346153846  0.019230769  0.942307692
     0.576923077  0.865384615  0.403846154
     0.961538462  0.942307692  0.173076923
     0.346153846  0.519230769  0.942307692
     0.269230769  0.903846154  0.788461538
     0.961538462  0.442307692  0.173076923
     0.807692308  0.211538462  0.865384615
     0.192307692  0.288461538  0.634615385
     0.500000000  0.750000000  0.250000000
     0.884615385  0.826923077  0.019230769
     0.576923077  0.365384615  0.403846154
     0.730769231  0.596153846  0.711538462
     0.115384615  0.673076923  0.480769231
     0.423076923  0.134615385  0.096153846
     0.653846154  0.980769231  0.557692308
     0.038461538  0.057692308  0.326923077
\end{verbatim}
\end{footnotesize}

\newpage
\bibliography{renormalized}